\documentclass[12pt,preprint]{aastex}

\shorttitle{Two possible circumbinary planets in NSVS\,14256825}
\shortauthors{Almeida, Jablonski \& Rodrigues}

\begin{document}

\title{Two possible circumbinary planets in the eclipsing post-common envelope system 
       NSVS\,14256825\footnote{Based on observations carried out at the 
       Observat\'orio do Pico dos Dias (OPD/LNA) in Brazil.}}

\author{L. A. Almeida, F. Jablonski and C. V. Rodrigues}
\affil{Divis\~ao de Astrof\'isica, Instituto Nacional de Pesquisas Espaciais,
    S\~ao Jos\'e dos Campos - SP, Brazil}

\email{leonardo@das.inpe.br}

\begin{abstract}
We present an analysis of eclipse timings of the post-common
envelope binary NSVS\,14256825, which is composed of an sdOB star and a dM star in a 
close orbit ($P_{\rm orb} = 0.110374$~days). High-speed photometry of this system was 
performed between July, 2010 and August, 2012. Ten new mid-eclipse times 
were analyzed together with all available eclipse times in the literature. We revisited the 
(O$-$C) diagram using a linear ephemeris and verified a clear orbital period variation.
On the assumption that these orbital period variations are caused by light travel time effects, 
the (O$-$C) diagram can be explained by the presence of two circumbinary bodies, even though
this explanation requires a longer baseline of observations to be fully tested.
The orbital periods of the best solution would be $P_c \sim 3.5$~years and $P_d \sim 6.9$~years. 
The corresponding projected semi-major axes would be $a_c\sin i_c \sim 1.9$~AU and 
$a_d \sin i_d \sim 2.9$~AU. 
The masses of the external bodies would be $M_c \sim 2.9~M_{\rm Jupiter}$ and 
$M_d \sim 8.1~M_{\rm Jupiter}$, if we assume their orbits are coplanar with the close binary.
Therefore NSVS\,14256825 might be composed of a close binary with two circumbinary 
planets, though the orbital period variations is still open to other interpretations.

\end{abstract}


\section{INTRODUCTION}
Planetary formation and evolution around binary systems have become important 
topics since the discovery of the first exoplanet around the binary pulsar 
PSR\,B1620-26 \citep{1993Natur.365.817B}. Theoretical studies 
have indicated that circumbinary planets can be formed and survive for a long 
time \citep{2004ApJ.609.1065M, 2006Icar..185.1Q}. Characterization of such planets 
in different evolutionary stages of the host binary is crucial to constrain and 
test the formation and evolution models. 

The common envelope (CE) phase in binary systems is dramatic for the planets survival. 
For a single star, \citet{2007ApJ.661.1192V} have pointed out 
that planets more massive than two Jupiter masses around a main sequence star of 
1 $M_{\odot}$ survive the planetary nebula stage down to orbital distances of 3 AU. 
The CE phase in binary systems is more complex than the nebular stage 
in single stars and its interaction with existing planets is still an open topic. 
Besides, a second generation of planets can be formed from a disk originated by 
the ejected envelope \citep{2009A&A.505.1221,2011AIPC.1331.56P}. 
Investigation of circumbinary planets in post-CE phase systems is 
fundamental to constrain observationally the minimum host binary-planet separation 
and to distinguish between planetary formation before and after the CE
phase.

To date, circumbinary planets have been discovered in seven eclipsing post-CE
binaries. All those discoveries were made using the eclipse timing 
variation technique. The main features of these planets are summarized
in Table~\ref{tab:known}, which also shows the results for NSVS\,14256825
presented in this paper.

NSVS\,14256825 (hereafter referred to as NSVS\,1425) is an eclipsing post-CE
binary and consists of an sdOB star plus a dM star with an orbital period of 
0.110374\,days \citep{2012MNRAS.423.478A}. It was discovered
using the Northern Sky Variability Survey \citep{Wozniak2004}. 
\citet{2012MNRAS.421.3238K} showed that the orbital period in NSVS\,1425 
is increasing at a rate of $\sim$\,$1.1 \times 10^{-10}~\rm s\,\,\rm s^{-1}$.
\citet{2012A&A.540A.8B} presented additional eclipse timings of NSVS\,1425 and
suggested from an analysis of the (O$-$C) diagram the presence of a
circumbinary planet of $\sim$\,$12~M_{\rm Jupiter}$.

In this study we present 10 new mid-eclipse times of NSVS\,1425 obtained 
between July, 2010 and August, 2012. We combined these data with previous 
measurements from the literature and performed a new orbital period variation 
analysis. In Section~\ref{observation} we describe our data as well as the reduction 
procedure. The methodology used to obtain the eclipse times and the
procedure to examine the orbital period variation are presented in 
Section~\ref{analysis}. In Section~\ref{discussion} we discuss our results.

\section{OBSERVATIONS AND DATA REDUCTION}\label{observation}

The observations of NSVS\,1425 are part of a program to search for eclipse timing 
variations in compact binaries. This project is being carried out using the facilities 
of the \textit{Observat\'orio do Pico dos Dias}/Brazil, which is operated by the 
\textit{Laborat\'{o}rio Nacional de Astrof\'{i}sica}. Photometric observations 
were performed using CCD cameras attached to the 0.6-m and 1.6-m telescopes. 
Typically $100$ bias frames and $30$ dome flat-field images were collected each 
night to correct systematic effects from the CCD data. The photometric data are 
summarized in Table \ref{table:1}.

\begin{figure}
 \centering
 \includegraphics[scale=0.8,angle=270]{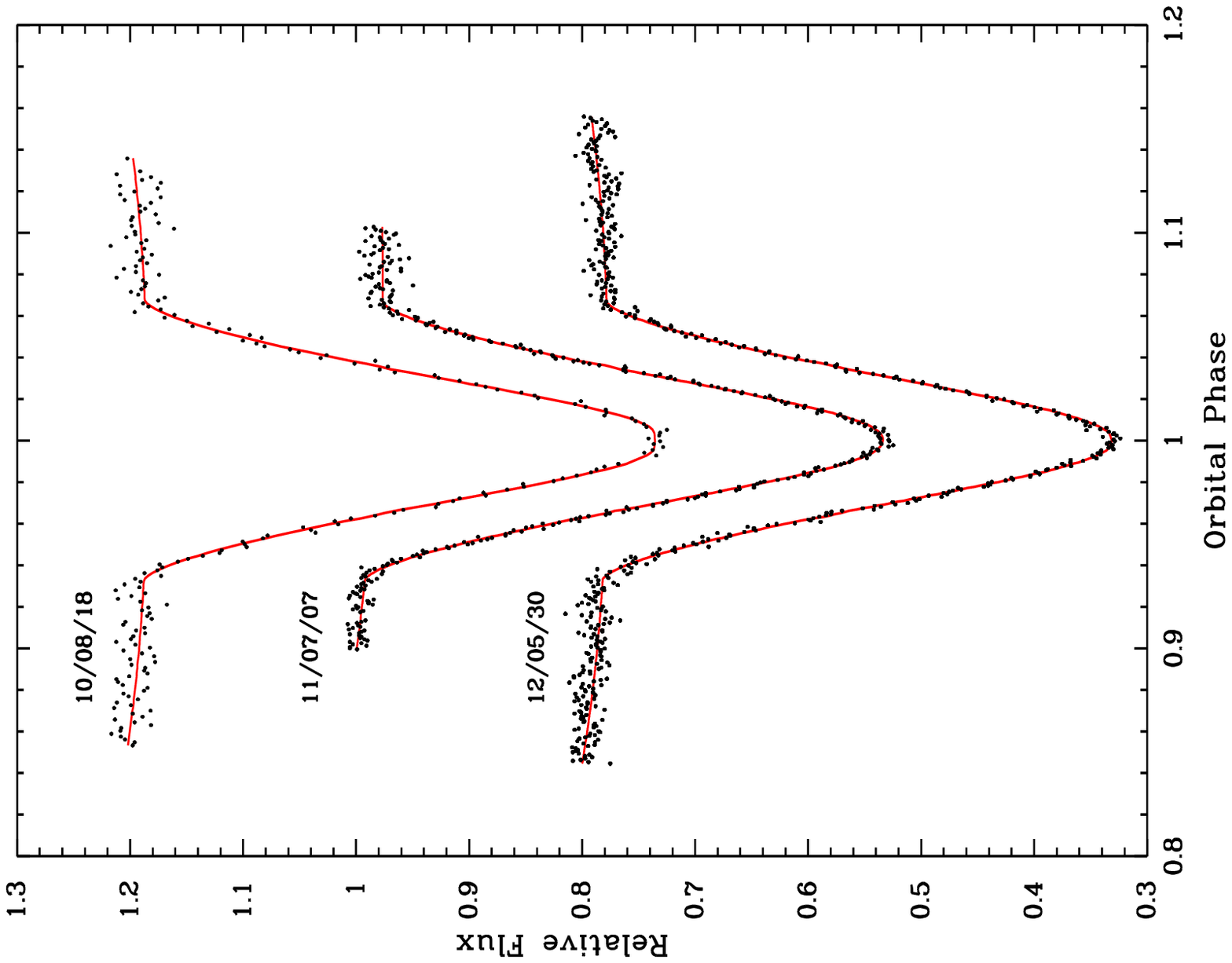}
 \caption{Primary eclipse of NSVS\,1425 folded on the binary orbital period.
          The solid line represents the best fitting performed with the 
          Wilson-Devinney code (see Section~\ref{analysis}). The upper
          and lower light curves were displaced vertically 0.2 units for
          better visualization.}
 \label{light_curve}
\end{figure}

The data reduction was performed using \textsc{iraf}\footnote{IRAF is distributed by the 
National Optical Astronomy Observatory, which is operated by the Association of 
Universities for Research in Astronomy (AURA) under cooperative agreement with the National 
Science Foundation.} tasks \citep{1993ASPC.52.173T}
and consists of subtracting a master median bias image from 
each program image, and dividing the result by a normalized flat-field frame. 
Differential photometry was used to obtain the flux ratio between 
the target and a field star of assumed constant flux. As the NSVS\,1425 field 
is not crowded, flux extraction was performed using 
aperture photometry. This procedure was repeated several times using different 
apertures and sky ring sizes to select the combination that provides the best
signal-to-noise ratio. Figure~\ref{light_curve} shows three normalized light curves 
folded on the NSVS\,1425 orbital period.  

\section{ANALYSIS AND RESULTS}\label{analysis}

\subsection{Eclipse fitting}
To obtain the mid-eclipse times for NSVS\,1425, we generated model light curves
using the Wilson-Devinney code (WDC -- \citealt{Wide1971}) and searched for the best fit to
the observed data. We used mode 2 of the WDC, which is appropriate to detached systems.  
The luminosity of each component was computed assuming stellar 
atmosphere radiation. The linear limb darkening coefficients, $x_i$, 
were used for both components. For the unfiltered light curves, $V$-band limb darkening 
was used. The ranges of the geometrical and physical parameters (e.g., inclination, 
radii, temperatures and masses) obtained by \citet{2012MNRAS.423.478A} for NSVS\,1425 
were adopted as the search intervals in the fit. 

A method similar to that described in \citet{2012MNRAS.423.478A} 
was used for the fitting procedure. The WDC was used as a ``function"~to 
be optimized by the genetic algorithm \textsc{pikaia} \citep{char1995}.
To measure the goodness of fit, we use the reduced $\chi_{\rm red}^2$ defined as

\begin{equation}
\chi^2_{\rm red}= \frac{1}{n}\sum_{j=1}^n\left(\frac{O_j-C_j}{\sigma_j}\right)^2,
\end{equation}

\noindent where $O_j$ are the observed points, $C_j$ are the corresponding models, 
$\sigma_j$ is the uncertainty at each point, and $n$ is the number of points. 
Figure~\ref{light_curve} shows three eclipses of NSVS\,1425 
and the corresponding best solutions. 
To establish realistic uncertainties, we used the solution obtained by 
\textsc{pikaia} as input to a Markov Chain Monte Carlo (MCMC) procedure \citep{Gilks1996} 
and examined the marginal 
posterior distribution of probability of the parameters. The mid-eclipse times and
corresponding uncertainties were obtained from the median value of the 
marginal distribution of the fitted times and the 1-$\sigma$ uncertainties from the 
corresponding 68\% area under the distribution. The results are presented in 
Table~\ref{timing}, together with previously published timings.

\subsection{Linear ephemeris}\label{ephemeris}

\begin{figure}
\centering
\includegraphics[scale=0.8,angle=-90]{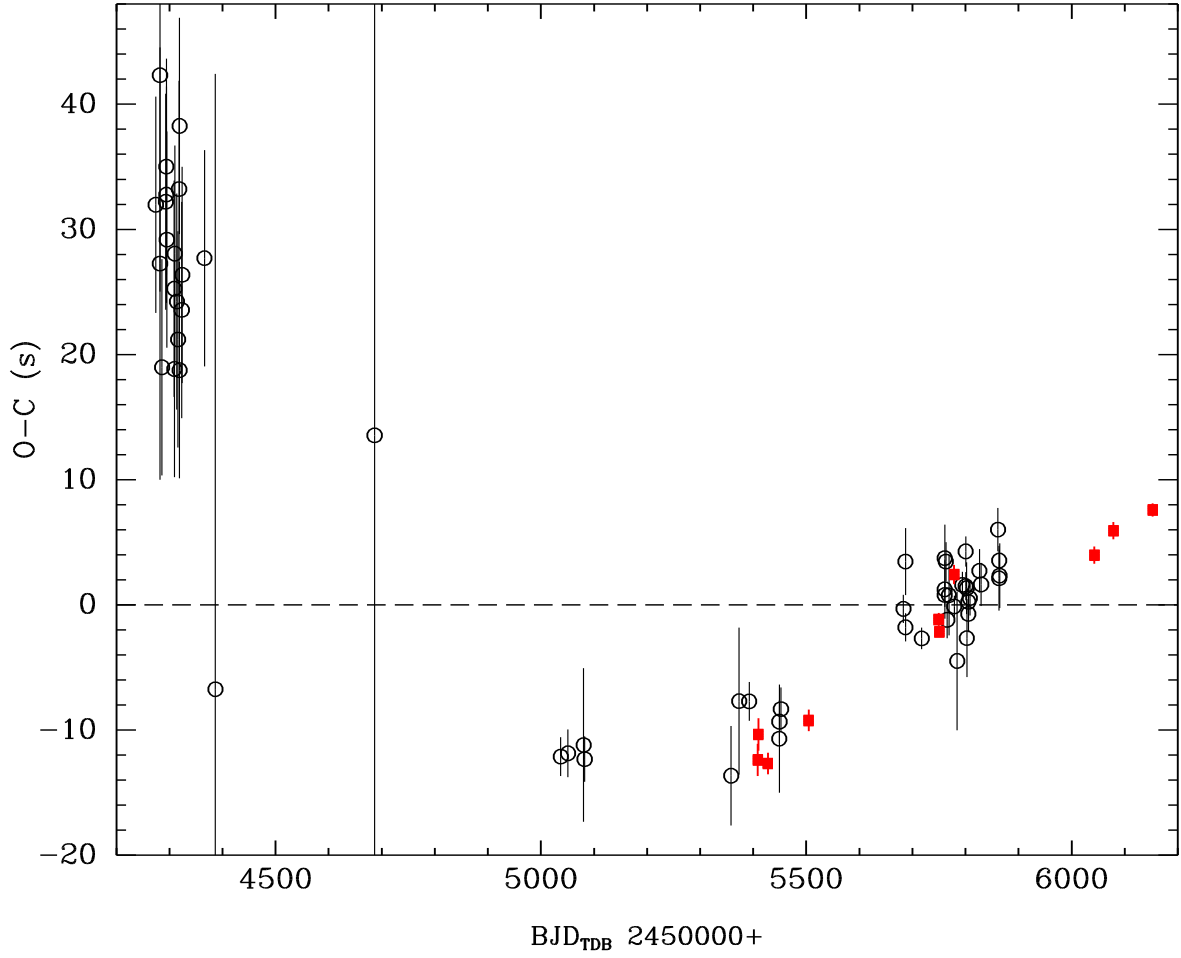}
\caption{(O$-$C) diagram of the eclipse timings of NSVS\,1425 made 
using Eq.~\ref{linear_ephem}. Our data are presented with full squares.}
\label{oc}
\end{figure}

To determine an ephemeris for the NSVS\,1425 orbital period, we analyzed 
our measurements together with all available eclipse times in the literature after 
converting them to barycentric dynamical time (TDB). Table~\ref{timing} shows all eclipse 
times available for NSVS\,1425. Fitting the data using a linear ephemeris, 
$T_{\rm min} = T_0 + E \times P_{\rm bin}$, we obtain

\begin{equation}
T_{\rm min} = {\rm TDB}\,2454274.2086(1) + 0.110374165(1) \times E,
\label{linear_ephem}
\end{equation}

\noindent where $T_{\rm min}$ are the predicted eclipse times, $T_{\rm 0}$ is a fiducial 
epoch, $E$ is the cycle count from $T_0$ and $P_{\rm bin}$ is the binary orbital period. 
The best fit yields a $\chi^2_{\rm red} \sim46$. The residuals of the observed times 
with respect to Eq.~\ref{linear_ephem} are shown in the (O$-$C) diagram of Fig.~\ref{oc}

\subsection{Eclipse timing variation}

Figure~\ref{oc} shows that a linear ephemeris is far from predicting
correctly the NSVS\,1425 eclipse times. The large value of $\chi^2_{\rm red}$ 
suggests the presence of additional signals in the (O$-$C) diagram.
One possible explanation is the light travel time (LTT) effect, which is explored in 
this paper.

The LTT effect shows up as a periodic variation in the observed eclipse 
times when the distance from the binary to the observer varies due to 
gravitational interaction between the inner binary and an external body
\citep{irwin1952}. To fit the NSVS\,1425 eclipse times taking this effect into
account, we used the following equation,

\begin{equation}
T_{\rm min} = T_0 + E\times P_{\rm bin} + \sum_{n=1}^n \tau_j,
\label{ephem_ltt}
\end{equation}

\noindent where

\begin{equation}
\tau_{\rm j} = \frac{z_j}{c}=K_{\rm j}\left[\frac{1-e^2_j}{1+e_j \cos f_j} \sin (f_j+\omega_j)\right]
\label{ltt}
\end{equation}

\noindent is the LTT effect. In the last equation, $K_j = a_j\sin i_j / c$ is the time
semi-amplitude, $e_j$ is the eccentricity, $\omega_j$ is the argument of periastron, 
and $f_j$ is the true anomaly. These parameters are relative to the orbit of the inner 
binary center of mass around the common center of mass consisting of the inner binary and 
of the $j-$th planet. The parameters $a_j$, $i_j$, and $c$ in the semi-amplitude equation 
are the semi-major axis, the inclination, and the speed of light, respectively. 
Notice that we do not consider mutual interaction between external bodies in this 
analysis.

\begin{figure}
\begin{center}
\includegraphics[angle=-90]{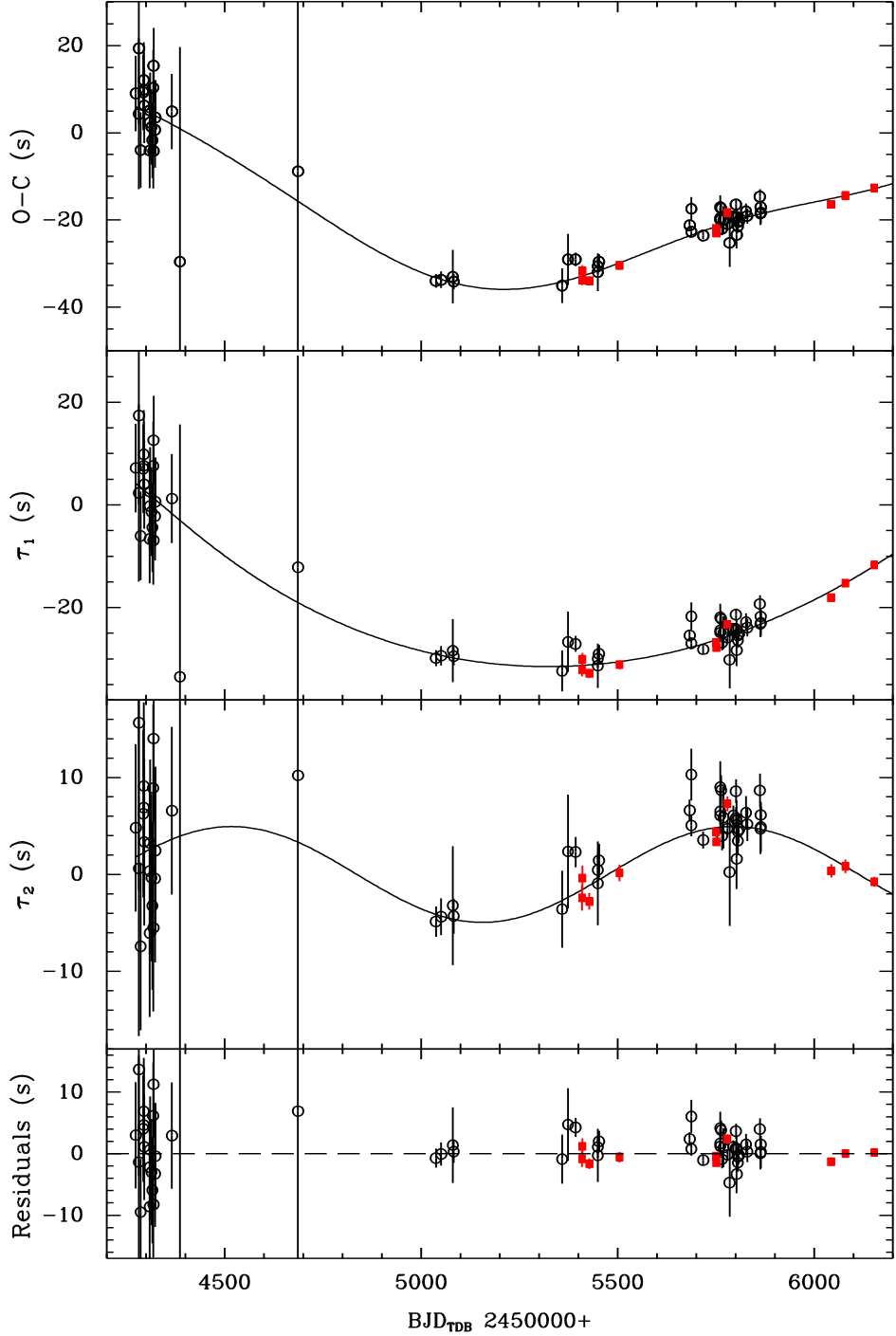}
\caption{The upper panel shows the (O$-$C) diagram of the eclipse times of NSVS\,1425
made with respect to the linear part of the ephemeris in Eq.~\ref{ephem_ltt}. Our data 
are presented with full squares and the solid line represents the best fit including 
the two LTT effects. The second and third panels display separately the two LTT 
effects ($\tau_1$ and $\tau_2$). The lower panel shows the residuals around the 
combined fit.}
\label{figo-c}
\end{center}
\end{figure}

Initially we fitted Eq.~\ref{ephem_ltt} to the data with only one LTT effect. 
The resulting $\chi_{\rm red}^2$ dropped to 6.8, but the new residuals showed evidences 
of another cyclic variation. Adding one more LTT effect in Eq.~\ref{ephem_ltt}, 
the resulting $\chi^2_{\rm red}$ improves to 1.85. The \textsc{pikaia} algorithm 
was used to search for the global optimal solution, followed by a MCMC procedure to sample 
the parameters of Eq.~\ref{ephem_ltt} around the best solution. 
Figure~\ref{figo-c} shows the resulting (O$-$C) diagram and Table~\ref{parameters} 
shows the best fit parameters with the associated $\pm 68\%$ uncertainties. 

\section{DISCUSSION AND CONCLUSION}
\label{discussion}

We revisited the orbital period variation of the post-CE binary NSVS\,1425 
adding the 10 new mid-eclipse times  obtained as described in previous sections. 
The complex orbital period variation, illustrated by the (O$-$C) diagram, can 
be mathematically described by the LTT effect of two circumbinary objects. 
The amplitudes of the LTT effects are  $\sim20$~s and $\sim5~s$. The associated 
orbital periods correspond to $\sim6.9$ and $\sim3.5$~years, respectively. This 
solution is a good description for the orbital period variation, as shown 
by Fig.~\ref{figo-c}. But it raises some concerns, which are discussed below.

The time baseline of NSVS\,1425 covers about 5.5~years. One of the two LTT 
effects has a period of $\sim7~$ years, larger than the baseline making the 
obtained solution less robust. Moreover, the early points have large errorbars 
and hence constrain less the LTT effect. In this regard, 
it is useful to recall the case of HW~Vir. It is a 
similar system, which also presents a complex (O$-$C) diagram. 
\citet{2003Obs.123.31K}, based on a dataset spread over almost 20~years, 
proposed the presence of a brown dwarf around the central binary. A few years later,  
a solution considering two objects was presented by \citet{2009AJ.137.3181L}. 
Recently, \citet{2012MNRAS.425.749H} claimed a still different solution, 
more stable dynamically. This illustrates how new data can 
change a LTT effect solution. Therefore, the  
LTT solution derived for NSVS\,1425 should be considered as a preliminary one.

We now discuss the implications of the presence of two circumbinary objects 
in NSVS\,1425.  Using the close
binary mass $M_{\rm bin}~=~0.528 M_{\odot}$ \citep{2012MNRAS.423.478A}, the lower
mass limit for the two circumbinary bodies are $M_c \sin i_c \sim 2.9~M_{\rm Jupiter}$
(inner body) and $M_d \sin i_d \sim 8.0~M_{\rm Jupiter}$ (external body). 
Assuming an orbital inclination of $82.5$ degrees \citep{2012MNRAS.423.478A} 
and coplanarity between the two external bodies and the inner 
binary, NSVS\,1425\,c and NSVS\,1425\,d would both be giant planets with 
$M_c \sim 2.9~M_{\rm Jupiter}$ and $M_d \sim 8.1~M_{\rm Jupiter}$.

Considering NSVS\,1425 with two circumbinary planets, this system would be 
the eighth post-CE system with planets and the fourth system with two planets 
(see Table~\ref{tab:known}). In such systems, there are two principal scenarios 
for planetary formation: 
(i) first generation planets formed in a circumbinary protoplanetary disk; and 
(ii) second generation planets originated from a disk formed by the ejected 
envelope \citep{2009A&A.505.1221,2011AIPC.1331.56P}.

In the first scenario, could the two circumbinary planets in NSVS\,1425 
survive the CE phase? \citet{2011MNRAS.411.1792B} estimated the orbital 
separation between the progenitor of an extreme horizontal branch (EHB) star (sdB or sdOB)
and a planet before the CE phase by the equation, 
$a_0 \simeq M_{\rm EHB} a_{\rm EHB} / M_{\rm pro}$, 
where $M_{\rm EHB}$ is the mass of the sdOB star, and $a_{\rm EHB}$ is the
present orbital separation of the planet. Assuming that the progenitor of 
the sdOB star in NSVS\,1425 had a mass of $M_{\rm pro} = 1.0\,M_{\odot}$, and neglecting 
accretion by the companion, the orbital separation of the planets before the CE
would be $a_{0c}\simeq0.8$ AU and $a_{0d}\simeq1.3$ AU. For a single star, 
\citet{2002MNRAS.336.449H} pointed out that the maximum radius at the tip of the EHB 
($R_{\rm EHB}$) is $\sim0.8$ AU. Hence, the inner planet is on the verge 
of being engulfed by the CE. Moreover, the tidal interaction causes a planet to 
spiral inwards if its orbital radius is smaller than $a_0 \lesssim 3\,R_{\rm EHB}$ 
\citep{2007ApJ.661.1192V}. 
Therefore the two circumbinary bodies in NSVS\,1425 would not survive the CE phase. 
On the other hand, \citet{2010NewAR.54.65T} showed that the CE size can be 
much reduced if the EHB star is part of a binary system, because once the secondary 
is engulfed by the envelope, the CE is totally ejected in only $\sim10^3$ days, 
stopping the envelope expansion. Therefore, the maximum radius of the CE is
around the initial distance between the close binary components.
Thus, if the close binary separation in NSVS\,1425 before the 
CE phase was $\lesssim 0.27$ AU, the two planets could survive the CE phase.

For the second scenario, the principal question to investigate is: was there
time enough after the CE phase to form giant planets? \citet{1999MNRAS.303.696K} 
showed that the typical time scale to form giant planets in protostellar disks 
is $\sim 10^6$ years. The lifetime of a binary in the EHB phase is  
$\sim 10^8$ years \citep{Dorman1993,2009ARAA.47.211H}. As the 
NSVS\,1425 primary star is in the post-EHB phase \citep{2012MNRAS.423.478A}, 
we conclude that there was time enough to form the two circumbinary planets after 
the CE phase. Therefore the second generation of planets is also a viable
scenario for NSVS\,1425.

Finally, among all known candidates to be circumbinary planets in post-CE 
systems, the inner planet in NSVS\,1425 has the minimum binary-planet separation,
$a_c \sin i_c \sim 1.9$\,AU. 

\section*{ACKNOWLEDGMENTS}
  This study was partially supported by CAPES (LAA), CNPq (CVR: 308005/2009-0), 
  and Fapesp (CVR: 2010/01584-8). We acknowledge the use of the SIMBAD 
  database, operated at CDS, Strasbourg, France; the NASA's Astrophysics Data System 
  Service; and the NASA's {\it SkyView} facility (http://skyview.gsfc.nasa.gov) 
  located at NASA Goddard Space Flight Center. 
  The authors acknowledge the referee, Dr. David Kilkenny, for his comments 
  and suggestions to improve this paper.

\begin{table}
\begin{center}
\caption{Circumbinary planets discovered in post-common envelope systems.}
\label{tab:known}
\begin{tabular}{lcccccr}
\hline
\hline
Name             &$P_{\rm orb}$&  $M_{\rm p} \sin i$&$a\sin i$&  e    & Spec. Type&  References \\
                 &  days       &$M_{\rm Jupiter}$   &  AU     &       &           &             \\
\hline
NSVS\,1425(AB)c  &  1276       &   2.9          & 1.9     & 0.0   & sdOB+dM    & this study \\
UZ For(AB)c      &  1917       &   7.7          & 2.8     & 0.05  &  DA+dM (Polar) & 1 \\
HU Aqr(AB)c      &  2226       &   4.5          & 3.32    & 0.11  &  DA+dM (Polar) & 2 \\ 
NSVS\,1425(AB)d  &  2506       &   8.0          & 2.9     & 0.52  & sdOB+dM    & this study \\                 
NN Ser(AB)c      &  2605       &   4.0          & 3.2     & 0.05  &  DA+dM     & 3  \\
NY Vir(AB)c      &  2900       &   2.3          & 3.3     & --    & sdB+dM     & 4  \\
RR Cae(AB)c      &  4346       &   4.2          & 5.3     & 0.0   &  DA+dM     & 5  \\
HW Vir(AB)c      &  4640       &   14.0         & 4.69    & 0.4   & sdB+dM     & 6  \\
HU Aqr(AB)d      &  5155       &   5.7          & 5.81    & 0.04  &  DA+dM (Polar) & 2  \\ 
NN Ser(AB)d      &  5571       &   6.71         & 5.32    & 0.22  &  DA+dM     & 3  \\
UZ For(AB)d      &  5844       &   6.3          & 5.9     & 0.04  &  DA+dM (Polar)  & 1  \\
DP Leo(AB)c      & 10227       &   6.05         & 8.18    & 0.39  &  DA+dM (Polar)  & 7 \\
\hline
\end{tabular}\\
\end{center}
\tablerefs{(1) \citet{2011MNRAS.416.2202P}; (2) \citet{2012MNRAS.420.3609H};
(3) \citet{2012MNRAS.425.749H}; (4) \citet{2012ApJ.745L.23Q}; 
(5) \citet{2012MNRAS.422L.24Q}; (6) \citet{2012AA.543A.138B};
(7) \citet{2011AA.526A.53B}.}
\end{table}

\begin{table}
\caption{Log of the photometric observations}   
\label{table:1}      
\centering                         
\begin{tabular}{l c c c c}        
\hline\hline                 
Date~~~~~ & $N$ & \ $t_{\rm exp}$(s) & Telescope & Filter  \\  
\hline                        
   2010 Jul 30  & 300  & 20 & 0.6-m & R$_C$  \\
   2010 Jul 31  & 450  & 20 & 0.6-m & R$_C$  \\
   2010 Aug 18  & 800  & 10 & 0.6-m & R$_C$ \\
   2010 Nov 20  &  350 & 20 & 0.6-m & I$_C$ \\
   2011 Jul 06  & 1255 & 2  & 1.6-m & I$_C$ \\
   2011 Jul 07  & 1300 & 1  & 1.6-m & Clear \\  
   2011 Aug 06  &  435 & 5  & 1.6-m & V     \\ 
   2012 Apr 24  & 1550 & 1.5& 0.6-m & Clear \\
   2012 May 30  &  600 & 3  & 0.6-m & Clear \\
   2012 Aug 12  & 1330 & 2  & 1.6-m & I$_C$ \\
\hline                                   
\end{tabular}
\end{table}

\begin{deluxetable}{lrrrrcrrrrr}
\tablewidth{0pt}
\tablecaption{Eclipse times for NSVS\,1425}
\tablehead{
\colhead{Cycle}          &
\colhead{Time (BJD--TDB)} & 
\colhead{O$-$C (s)}            & 
\colhead{Eclipse}        & 
\colhead{Ref.}}
\startdata
 1      &  2454274.2088(1)   &   9.0    & I     & 1  \\
 72     &  2454282.1559(2)   &  19.4    & I     & 1 \\
 73     &  2454282.2661(2)   &   4.3    & I     & 1 \\
108     &  2454286.1291(1)   &  -4.0    & I     & 1  \\
172     &  2454293.1932(1)   &   9.2    & I     & 1  \\
180     &  2454294.0762(1)   &   9.8    & I     & 1  \\
181     &  2454294.1866(1)   &  12.1    & I     & 1  \\
190     &  2454295.1799(1)   &   6.2    & I     & 1  \\
316     &  2454309.0870(1)   &   2.3    & I     & 1  \\
317     &  2454309.1973(1)   &  -4.1    & I     & 1  \\
325     &  2454310.0804(1)   &   5.1    & I     & 1  \\
362     &  2454314.1642(1)   &   1.3    & I     & 1  \\
380     &  2454316.1509(1)   &  -1.7    & I     & 1  \\
397     &  2454318.0274(1)   &  10.3    & I     & 1  \\
406     &  2454319.0206(1)   &  -4.2    & I     & 1  \\
407     &  2454319.1312(1)   &  15.4    & I     & 1  \\
443     &  2454323.1045(1)   &   0.7    & I     & 1  \\
452     &  2454324.0979(1)   &   3.5    & I     & 1  \\
832     &  2454366.0401(1)   &   4.9    & I     & 1  \\
1018    &  2454386.5693(6)   & -29.6    & I     & 2  \\
3737    &  2454686.6769(5)   &  -8.8    & I     & 2  \\
6914    &  2455037.33534(2)  & -34.0    & I     & 2  \\
7037    &  2455050.91137(2)  & -33.7    & I     & 2  \\
7304    &  2455080.38128(7)  & -33.0    & I     & 2  \\
7322    &  2455082.36800(2)  & -34.1    & I     & 2  \\
9823.5  &  2455358.46897(5)  & -35.0    & II    & 2  \\
9959    &  2455373.42474(7)  & -29.1    & I     & 2  \\
10131   &  2455392.40910(2)  & -29.0    & I     & 2  \\
10279   &  2455408.74442(2)  & -33.7    & I     & this study        \\
10287   &  2455409.62744(2)  & -31.7    & I     & this study        \\
10451   &  2455427.72877(1)  & -34.0    & I     & this study        \\
10646   &  2455449.25176(5)  & -31.9    & I     & 3  \\
10647   &  2455449.36215(2)  & -30.6    & I     & 3  \\
10673   &  2455452.23189(2)  & -29.6    & I     & 3  \\
11146.5 &  2455504.49405(1)  &  -30.4   & II    & this study       \\
12763   &  2455682.91400(2)  &  -21.2   & I     & 2 \\
12799   &  2455686.88745(2)  &  -22.7   & I     & 2 \\
12799.5 &  2455686.94270(3)  &  -17.4   & II    & 2 \\
 13077  &  2455717.57146(1)  &  -23.5   & I     & 3 \\
 13368  &  2455749.690361(6) &  -22.0   & I     & this study \\    
 13377  &  2455750.683717(4) &  -23.0   & I     & this study \\    
 13629  &  2455778.498061(9) &  -18.3   & I     & this study \\    
 13469  &  2455760.83818(3)  &  -19.6   & I     & 2 \\
 13469.5&  2455760.89340(3)  &  -17.1   & II    & 2 \\
 13470  &  2455760.94855(1)  &  -20.0   & I     & 2 \\
 13488  &  2455762.93532(2)  &  -17.3   & I     & 2 \\  
 13511  &  2455765.47387(2)  &  -22.0   & I     & 2 \\  
 13542  &  2455768.89549(4)  &  -20.0   & I     & 2 \\  
 13632  &  2455778.82915(1)  &  -20.9   & I     & 2 \\ 
 13682  &  2455784.34781(7)  &  -25.2   & I     & 2 \\   
 13768  &  2455793.84006(1)  &  -19.2   & I     & 2 \\ 
 13827  &  2455800.35217(2)  &  -16.5   & I     & 2 \\ 
 13828  &  2455800.46251(2)  &  -19.2   & I     & 2 \\
 13845  &  2455802.33887(5)  &  -19.4   & I     & 2 \\
 13846  &  2455802.44920(4)  &  -23.4   & I     & 2 \\
 13872  &  2455805.31896(2)  &  -20.5   & I     & 2 \\
 13873  &  2455805.42932(2)  &  -21.5   & I     & 2 \\
 13899  &  2455808.29907(2)  &  -20.2   & I     & 2 \\
 14062  &  2455826.29008(2)  &  -18.0   & I     & 3 \\
 14089  &  2455829.27017(2)  &  -19.1   & I     & 3 \\
 14379  &  2455861.27873(2)  &  -14.6   & I     & 3 \\ 
 14397  &  2455863.26542(3)  &  -18.5   & I     & 3 \\
 14400  &  2455863.59656(1)  &  -17.1   & I     & 2 \\
 14406  &  2455864.25879(3)  &  -18.3   & I     & 3 \\
 16024  &  2456042.844216(4) &  -16.4   & I     & this study \\    
 16350  &  2456078.826240(8) &  -14.4   & I     & this study \\    
 17019  &  2456152.666554(6) &  -12.6   & I     & this study \\    
\enddata
\tablerefs{
(1) \citet{Wils2007}; (2) \citet{2012AA.543A.138B}; 
(3)\citet{2012MNRAS.421.3238K}.}
\label{timing}      
\end{deluxetable}

\begin{table}
\begin{center}
\caption{Parameters for the linear plus two-LTT ephemeris of NSVS\,1425.\label{parameters}} 
\begin{tabular}{l c r }        
\tableline\tableline                   
           & \small Linear ephemeris  &  \\
\tableline
Parameter  & Value & Unit   \\    
\tableline                        
$P_{\rm bin}$  & 0.1103741681(5)   & days  \\
$T_0$    & 2454274.20874(4)  & BJD(TDB)  \\
\tableline
   & \small $\tau_1$ term   &  \\
\tableline
Parameter  & Value & Unit   \\    
\tableline                        
$P$             &  $6.86\pm0.45$                   & years   \\
$T$             &  $2456643\pm110$                  & BJD(TDB) \\
$a_{\rm bin} \sin i$ &  $0.042\pm0.006$            & AU      \\
$e$             &  $0.52\pm0.08$                   &         \\
$\omega$        &  $98\pm9$                      & degrees \\
$f(M)$          & $(1.6\pm1.1) \times 10^{-6}$     & $M_{\odot}$ \\ 
$M \sin i$      & $8.0\pm1.5$                      & $M_{\rm Jupiter}$  \\ 
$a \sin i$      & $2.9\pm0.7$                      & AU \\
\tableline
   &\small $\tau_2$ term  &  \\
\tableline
Parameter  & Value & Unit   \\    
\tableline                        
$P$            & $3.49\pm0.38$                   & years  \\
$T$            & $2455515\pm95$                  & BJD(TDB)   \\
$a_{\rm bin}\sin i$ & $0.0099\pm0.0006$          & AU     \\
$e$            & $0.00\pm0.08$                   &        \\
$\omega$       & $11\pm8$                      & degrees \\
$f(M)$         & $(8.0\pm4.0) \times 10^{-8}$    & $M_{\odot}$  \\ 
$M \sin i$     & $2.9\pm0.4$                     & $M_{\rm Jupiter}$  \\ 
$a \sin i$     & $1.9\pm0.8$                     & AU \\
\tableline
$\chi^2_{\rm red}$&  1.85                        &          \\
\tableline                         
\end{tabular}%
\end{center}
\end{table}

\end{document}